\documentclass{JHEP}
\usepackage{amsmath,amsfonts,bbm,euscript, amssymb}
\usepackage{cite}
\usepackage{epsfig}
\usepackage{graphicx}
\usepackage{psfrag}
\usepackage{epsf}
\input{epsf.sty}

\def\be{\begin{equation}}
\def\ee{\end{equation}}
\def\baray{\begin{eqnarray}}
\def\earay{\end{eqnarray}}
\def\ba{\begin{eqnarray}}
\def\ea{\end{eqnarray}}

\title{A Note on the Quantum Creation of Universes}
\author{Saswat Sarangi $^1$ and S.-H. Henry Tye $^2$\\
 $^1$ Institute of Strings, Cosmology and Astroparticle Physics
Department of Physics, Columbia University, New York, NY 10027\\
$^2$ Laboratory for Elementary Particle Physics, Cornell
University, Ithaca, NY 14853 \\
E-mail:\email{sash@phys.columbia.edu},
\email{tye@lepp.cornell.edu}}

\date{\today}

\abstract{ We elucidate the nature of the correction to the Hartle-Hawking wavefunction
presented in hep-th/$0505104$ and hep-th/$0406107$. The correction comes from the quantum fluctuation of the metric that spontaneously breaks the classical deSitter symmetry. 
This converts the tunneling from nothing to a deSitter-like universe via a $S^{4}$ instantion to that via a barrel instanton, which is bounded from below. Its generalization to 10 dimensional spacetime allows us to find the preferred sites in the stringy cosmic landscape. 
We comment on how some of the problems of the Hartle-Hawking 
wavefunction are avoided with the new modified wavefunction of the universe, when applied to the 
spontaneous creation of an inflationary universe.
We also summarize our arguments on the validity of the Hartle-Hawking wavefunction in the minisuperspace approximation, 
as opposed to the WKB formula suggested by Linde and Vilenkin. 
}

\begin{document}

\section{Introduction and Summary}

In recent years, it has become clear that the vacuum structure in superstring theory
is very rich. The number of metastable discrete vacua is estimated to be somewhere between $10^{300}$ to $\infty$. Different vacau have different forces and matter content. This is the cosmic landscape. With these many possible vacuum states, one is led to ask why we end up where we are, in a particular vacuum. One may choose to believe in the (strong) anthropic principle and so avoid addressing this very deep and fundamental question. Alternatively, one may try to find a physics reasoning to gain a better understanding of the questions ``why and how''.
 
In an attempt to address this question, we start with Vilenkin's beautiful idea of tunneling from nothing
\cite{Vilenkin:1982de}, or equivalently, the no-boundary wave function of Hartle and Hawking \cite{Hartle:1983ai}. Here, nothing means no classical space-time. Generalizing to 10-dimensional spacetime, one can calculate the tunneling probability from nothing to every state in the cosmic landscape in superstring theory. Our universe should begin with the state that has the maximum (or close to the maximum) tunneling probability
from nothing. Hopefully, one finds that our inflationary universe is among the ones preferred. If that inflationary universe then evolves to our today's universe, one would have a physical understanding why we end up where we are. As a byproduct, we should also understand why our observable universe has a 3+1 dimensional observable space-time. Unfortunately, the Hartle-Hawking (HH) wavefunction of the universe does not seem to give the answer we are looking for. 

\begin{figure}
\begin{center}
\epsfig{file=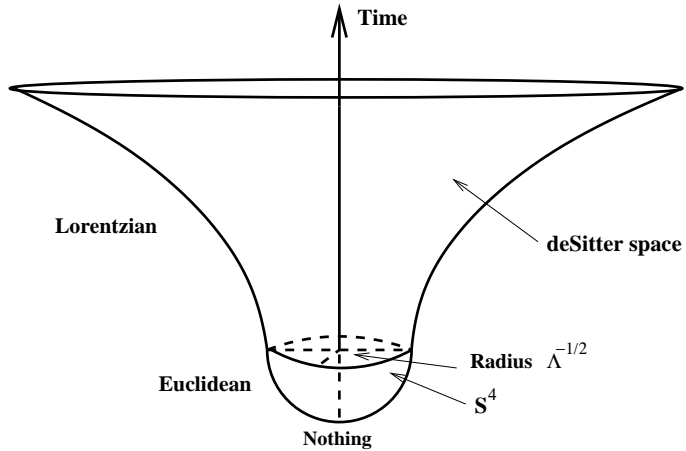, width=9cm}
\vspace{0.1in}
\caption{The creation of a closed deSitter universe  from nothing.}
\label{fig1}
\end{center}
\end{figure}

Our proposal\cite{Firouzjahi:2004mx,Sarangi:2005cs} introduces a correction term to the HH wavefucntion for a closed universe. Since tunneling is a quantum process, consistency requires the inclusion of quantum corrections to the resulting universe as well.
The correction term is simply the result of including this quantum correction.
By going a little outside the mini-superspace approximation, this correction remedies the problem faced by the HH wavefunction. We can also see how this correction 
emerges in an explicit calculation. Although the improved wavefunction and its application 
to the landscape concerns 10-dimensional spacetime in string theory, we shall simplify the discussion here by focusing on 4-dimensional Einstein theory.

In the usual evaluation of the quantum one-loop correction to the global Euclidean deSitter space, 
one constrains the quantum correction to obey the symmetry of the $S^4$ instanton (i.e., the 
deSitter $SO(5)$ rotational symmetry). As a consequence of this choice of renormalization condition, the only possible correction is a renormalization of the cosmological constant (and/or of the curvature term) in pure de-Sitter space. However, there is no apriori reason for this. In fact, the deSitter space has a horizon and so an effective Hawking temperature.  So, quantum fluctuations naturally introduce tensor modes as metric perturbations (a well-known fact in the inflationary universe). Here, such modes play the role of radiation, and their presence brings us outside the mini-superspace approximation. 
Taking into account its back reaction on the metric, this perturbation breaks the pure deSitter symmetry.
Although mathematically one can get a pure quantum deSitter space, it is more natural physically to choose the renormailzation condition  so that the classical pure deSitter symmetry is broken by the quantum effect. Our key position is that there is no pure quantum deSitter space in this physical sense.
As a result, there is no tunneling to a pure deSitter universe. Our calculation shows that this spontaneous breaking of the deSitter symmetry converts 
the $S^4$ instanton to a barrel instanton \cite{Sarangi:2005cs}. 

The quantum fluctuation appears as radiation in Lorentzian metric. Integrating out the metric perturbative modes in Euclidean metric, we find a new term that corrects the HH wavefunction.
Instead of the $S^4$ instanton, we now end up with a barrel instanton. In terms of the tunneling probability from nothing to a quantum-corrected deSitter universe with cosmological constant 
$\Lambda$, the radiation introduces a new term to the Euclidean action,
\ba
\label{Big}
P  \sim \exp(-S_{E})= \exp({3\pi/G \Lambda})  \rightarrow  \exp \left(\frac{3\pi}{G\Lambda} - \frac{b M_s^4}{\Lambda^2} \right)
\ea
where $G=M_{Planck}^{-2}$ is the Newton's constant, $b$ is a numerical constant and $M_{s}$ is the ultra-violet cut-off. This cut-off breaks the conformal property. In string theory, this UV cut-off is provided by the string scale. Since we are considering a closed universe, there is no modes with wavelength longer than the size of the closed universe. So there is a natural infrared cut-off in the  wave number 
of the metric perturbative modes, namely, the finite size $\sim \Lambda^{-1/2}$. This leads to the 
$\Lambda^{-2}$ factor. Besides the tensor modes, all fields with mass smaller than $M_{s}$ will contribute, so $b$ depends on the specific matters fields present. 
Maximizing the tunneling probability, one finds $\Lambda_{max} = 2bM_s^4/3 \pi M_{Planck}^2$. Typically the string scale is below the Planck scale, and the logarithm of $b$ is of order unity, so $\Lambda_{max}$ is much smaller than the Planck scale. In string theory, different sites in the landscape with 3 large spatial dimensions will have different values of $b$. Recall that, without the new term (i.e., the original HH wavefunction), $\Lambda=0$ would have been preferred. This is due to the unboundedness of the Euclidean action $S_{E}$. On the other hand, the improved barrel instanton action is bounded from below. 

In Ref\cite{Firouzjahi:2004mx}, we use the concept of decoherence to argue for the existence of such a correction. In Ref\cite{Sarangi:2005cs}, the calculation of the barrel instanton is motivated along this line of thinking. The metric perturbation that leads to the radiation term occurs over a Hubble time and so is effectively non-local. This radiation acts as the environment to the cosmic scale factor $a$ (the system) and thus suppresses the tunneling, i.e., the spontaneous creation of the universe from nothing. 
Based on the well-understood idea of decoherence in quantum measurement \cite{Caldeira:1982uj}, this is precisely what one expects. As $\Lambda$ decreases, there are more modes contributing to the
environment, hence further suppression. Also, the presence of more matter fields tends to increase the suppression of the tunneling probability as well, so $b$ measures the degrees of freedom. Extending the wavefunction to 10-dimensions is straightforward. In the case where 6 dimensions are compactified, the compactification properties are encoded in $G$, $\Lambda$ and $b$ in Eq(\ref{Big}). 

One can now use this improved wavefunction (i.e., its generalization to 10-dimensions) to study the spontaneous creation of universes and find out why some vacua are preferred over others.
Among known string vacua (such as any supersymmetric vacuum, a 10-dimensional deSitter-like  vacuum etc.), a 3+1 dimensional inflationary universe not unlike ours seems to be preferred. Here we
have in mind, for example, the brane inflationary scenario in the brane world in string 
theory \cite{Firouzjahi:2004mx,Dvali:1998pa,Kachru:2003sx}. 
It is important to note that, instead of our universe today (with a tiny cosmological constant), the tunneling probability picks an inflationary universe, which can then evolve to our universe today. We refer to this as SOUP, the selection of the original universe principle.
This preliminary result is attractive enough to deserve further analysis. 
In fact, the barrel instanton seems to have distinct cosmological signatures that, at least in principle, may be detected in the cosmic microwave background radiation \cite{Sarangi:XX}. 

It is obviously important to understand the precise meaning and the properties of what we refer to as ``nothing'', with no classical spacetime. Since spacetime appears to be a derived quantity in string theory, we expect ``nothing''  to be fully meaningful in string theory. Just like quantum field theory provided us a good understanding of the vacuum in the past century, string theory should provide us a good understanding of the full quantum nature of ``nothing''. Conceptually, this may be the ultimate challenge in string theory. Here, the tunneling probability is calculated in the semi-classical approximation. Fortunately, the vacua with high probabilities are precisely the ones with the effective cosmological constant small compared to the Planck scale, so the semi-classical approximation should be valid here. So we are hopeful that the improved wavefunction in its semi-classical approximation will allow us to select the preferred site in the cosmic landscape. 

As mentioned earlier, the original HH wavefunction seems to prefer a zero cosmological constant $\Lambda$, corresponding to an infinite size universe ($a \sim {\Lambda}^{-1/2}$, thus skipping the big bang era). This clearly disagrees with observation, is counter-intuitive, and, some believe \cite{Coleman:1988tj,Fischler:1990se}, inconsistent. 
This problem motivated Linde and Vilenkin to seek an alternative tunneling formula or 
wavefunction\cite{Linde:1984mx}, which we call the LV wavefunction.
Since our new wavefunction is predicated on the HH wavefunction, not on the LV wavefunction, 
it is important for us to explain the correctness of the HH wavefunction as the zeroth order approximation. This is one of the purposes of this note.

In quantum field theory, the tunneling probability from one state to another state is well-understood.
One may use the instanton method, where the tunneling probability $P$ is given by the Euclidean action $S_{E}$, namely $P \simeq e^{-S_{E}}$ as in HH wavefunction. 
Alternatively, one may use the WKB method in the Wheeler-DeWitt formalism, where the tunneling amplitude is given in turns of the integral of the momentum across the barrier. It is understood that typical tunneling process is exponentially suppressed, so the negative sign is chosen, as in the LV wavefunction, so $P \simeq e^{ - 2 \int |p| dr}$. In quantum field theory, these 2 approaches yield the same answer. However, when applied to gravity, the Euclidean action can be unbounded from below. In fact, for tunneling from nothing to a closed deSitter universe, $S_{E}= -3\pi/G \Lambda$ where $\Lambda$ is the cosmological constant and $G$ is the Newton constant. In the instanton method, this naively seems to lead to an exponentially large tunneling probability for the spontaneous creation of a universe as $\Lambda \rightarrow 0$, while the LV wavefunction strongly disfavors this limit. This leads to a puzzle, as illustrated in Fig. \ref{fig1}. This is the 20 years old debate on which sign is correct \cite{Vilenkin:1998rp}.
 
Our proposal introduces a new term to the HH wavefunction,
implying that the HH wavefunction is the correct starting point, not the alternative LV version. Furthermore, the improved wavefunction no longer suffers from the above-mentioned problems that originally led to the consideration of alternatives. Looking at the WKB approximation, we see that the other WKB solution, namely, $P \simeq e^{ + 2 \int |p| dr}$ is the correct choice for the tunneling processes here.

\begin{figure}
\begin{center}
\epsfig{file=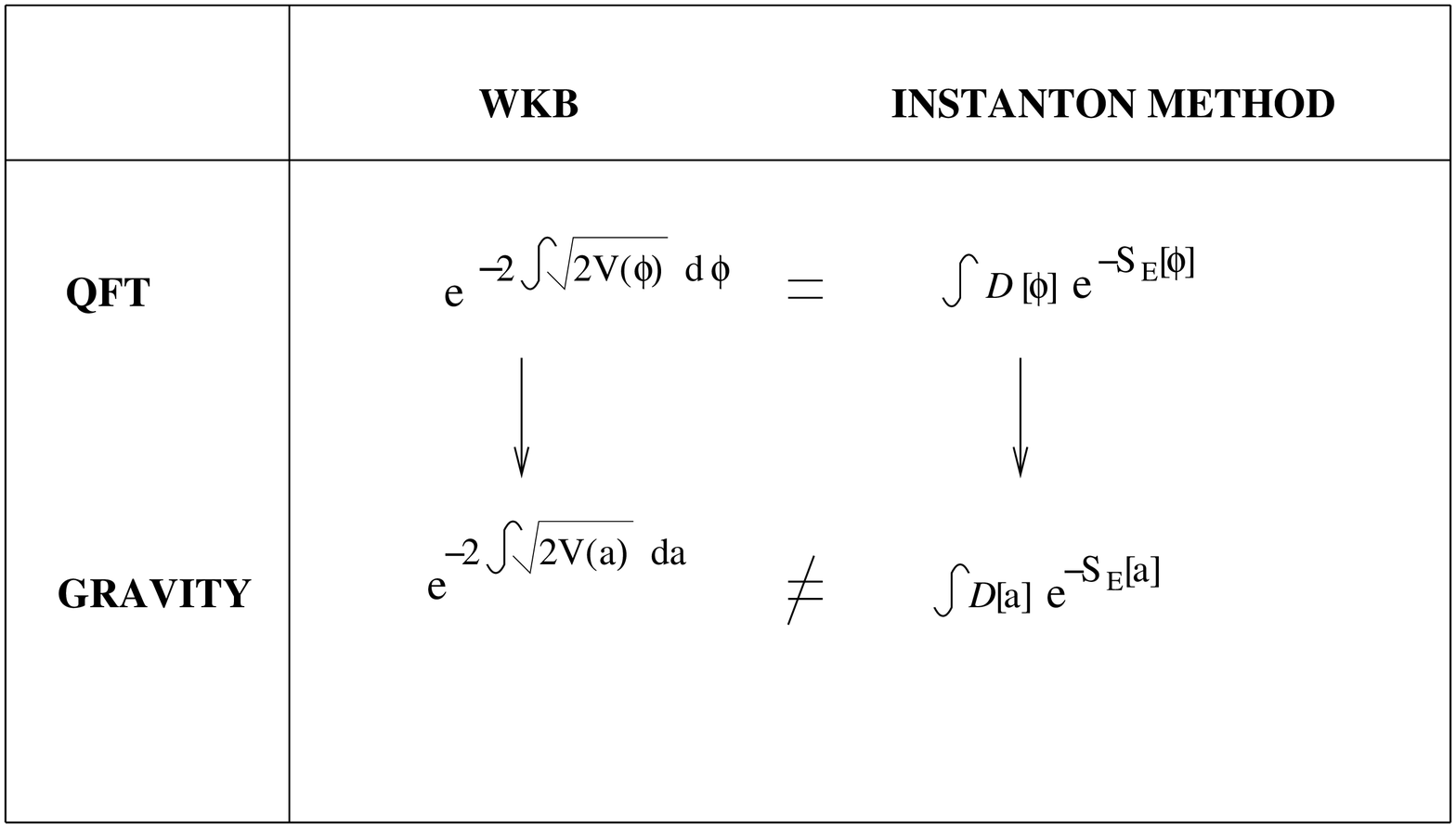, width=14cm}
\vspace{0.1in}
\caption{Debate about the correct sign in the exponent in the tunneling probability from nothing to a closed deSitter (inflationary) universe. The WKB and the instanton approaches agree in quantum field theories. However, due to the negative sign of $S_{E}[a]$ (and its unboundedness from below), they seem to disagree in the quantum creation of the universe. We argue that the instanton approach, which in the zero-order approximation corresponds to the Hartle-Hawking wavefunction, is correct. That is, one should take the other (i.e., the plus sign term) WKB solution in this particular application.}
\label{fig1}
\end{center}
\end{figure}

In this note we like to further clarify the modified wavefunction and collect all the arguments, some old and some new, for the HH version and against the LV version : 


$\bullet$ In the instanton formulation, Lorentzian time is rotated to Euclidean time via 
$t \rightarrow -i\tau$. For quantum fields, this leads to a consistent formulation, where a realistic scalar field Euclidean action will be bounded from below. 
In the LV version, one has  to rotate time the other way, i.e.,  $t \rightarrow i\tau$. In this case, the scalar field Euclidean action will be unbounded from below. In pure gravity, such an unusual Wick rotation may be acceptable. However, any realistic stringy vacuum or gravity theory will include scalar fields (say, the moduli in string vacua and the inflaton field in an inflationary model). So consistency of the scalar field will dictate the correct Wick rotation, which chooses the HH version. The alternative will cause catastrophic behavior in the scalar fields. This issue has been discussed by Rubakov and others \cite{Rubakov:1984bh}.

 $\bullet$ Recently, a topological string theory calculation yields the HH wavefunction, not the LV wavefunction. In Ref\cite{Ooguri:2005vr}, Ooguri, Vafa and Verlinde argue that the topological string partition function can be interpreted as an exact HH wavefunction of the universe in the minisuperspace sector of the physical superstring theory, including all loop quantum corrections. Although they derived the wavefunction of the universe in a cosmology with a negative cosmological
constant, it can be seen as the derivation of a Hartle-Hawking like result in string theory. The geometry considered is a Euclideanized version of $AdS^2 \times S^2 \times M$, where $M$ is a Calabi-Yau 6-fold; the result includes all loop corrections. Their proposal leads to a conceptual understanding of the observation that the black hole entropy is the square of the topological string wavefunction. 
The wavefunction thus obtained prefers a universe with vanishing cosmological constant.
Although their result is for topological string theory in the minisuperspace approximation, the preference for HH over LV wavefunction will most likely survive in the physical superstring theory.

 $\bullet$ In Ref.\cite{Sarangi:2005cs}, we use decoherence to find the correction term to the HH wavefunction. Going beyond the mini-superspace, we introduce fluctuations around the background metric as the environment where the scale factor $a$ is the system. This may be considered as a quantum ``back-reaction improved'' 1-loop effect. The tunneling is then suppressed by its interaction with the environment. This is expected. The tunneling is from nothing to a universe with size $a \simeq H^{-1}\simeq 1/\sqrt{\Lambda}$. As $\Lambda \rightarrow 0$, the resulting universe becomes infinite. The tunneling to such a super-macroscopic universe must be suppressed. Indeed, this suppression is the decoherence effect. 
 
Note that the quantum fluctuation considered here has the same origin as the tensor perturbation studied in an inflationary universe. There, one is interested in the tensor mode with wave number
$k =aH$ as it leaves the horizon. Here we are interested in all modes. It is the sum over all modes that requires the introduction of an ultra-violet cut-off. After tunneling, their contribution behaves as a radiation term,
which fixes the sign of this correction to the $S^{4}$ instanton.   
If we apply the decoherence effect to the LV wavefunction, this environment actually enhances the tunneling, contrary to our expectation from decoherence. Hence the decoherence interpretation selects the HH wavefunction as the correct wavefunction at zeroth order.
  
$\bullet$ One may think that the issue with the HH wavefunction is intimately related to the unboundedness in the gravity theory. Finding another way to derive tunneling probabilities in Euclidean gravity theory should help to clarify some of the issues.
In the stochastic formulation for an inflationary universe developed by Starobinsky and others
\cite{Starobinsky:1986fx,Linde:2005ht},
one can derive the tunneling rate from one $\Lambda$ to another $\Lambda$; it is again given by the instanton formalism, not the LV formalism. We shall elaborate on this approach and motivate a correction term (similar to the one due to decoherence effect in path integral approach), shedding further insight to the underlying physics.

$\bullet$ In usual field theory, the tunneling probability is dominated by the most symmetric instanton.
Here, it would have been the $S^{4}$ instanton. Instead, as argued in Ref\cite{Coleman:1988tj,Fischler:1990se}, this is not the case here, leading to the consideration of other complicated instantons, such as wormholes and a chain of $S^{4}$ bubbles. This is a consequence of the unboundedness of the Euclidean action of the $S^{4}$ instantoin.
Naively, the LV wavefunction avoids this problem. However, the LV wavefunction prefers large $\Lambda$, in a region where the semi-classical approximation breaks down anyway. 
In contrast, the proposed modified Euclidean action has a lower bound, and so offers a chance to avoid the disaster encountered by the $S^{4}$ instanton. We like to point out that even in quantum theory, the most symmetric instanton will be modified in the presence of an environment. Here, the $S^{4}$ instanton would be recovered in the limit where the cut-off (i.e., superstring scale) approaches zero.

Once we adopt the improved HH wavefunction, we find that it can be readily applied to the selection of the original (initial) universe in the cosmic string landscape. The original problem associated with the HH wavefunction (i.e., universes with a zero cosmological constant are preferred) no longer holds. Instead, an inflationary universe seems to be preferred. In particular, a 4-dimensional inflationary universe not unlike ours seems to be a prime candidate with relatively large tunneling probability.
If instead we blindly apply the correction to the LV wavefunction, we find that it does not yield anything new, rather, a large cosmological constant universe will again be preferred. In the large cosmological constant case, semi-classical approximation breaks down and this whole approach becomes moot.  

It has been argued by DeWitt and others \cite{DeWitt:1967yk} that, to avoid a singularity at $a=0$, the wavefunction there must vanish, i.e., $\Psi(a=0)=0$. This argument becomes more relevant when one considers tunneling from nothing to any stringy vacua, where typically the scale factor ${\bf a}$ is a multi-dimensional vector. A non-vanishing $\Psi$ at ${\bf a}={\bf 0}$  will introduce an arbitrary initial condition, which violates the basic notion of nothing.  Any choice of initial condition other than nothing will bias the selection of preferred vacua, and furthermore raise the question of why that particular choice of initial condition.
With $\Psi(a=0)=0$, only the HH wavefunction makes sense. To see why this happens, we refer to Fig.($3$). In the under-the-barrier region the Wheeler-DeWitt equation yields an exponentially growing and an exponentially decaying mode as its solutions. Clearly, the boundary condition that the wavefunction should vanish at $a = 0$ requires the presence of the growing mode. The growing mode attains a maximum value that is the exponential of the inverse of $\Lambda$. One can see why this is so. The  base of the under-the-barrier region is proportional to $\sqrt{3/\Lambda}$. Smaller the value of $\Lambda$, larger the under-the-barrier region and larger the maximum value that the growing mode attains. This is precisely the
HH result. The HH wavefunction goes as $\exp(3\pi/G\Lambda)$. Hence, the condition $\Psi(a=0)=0$ is consistent only with the HH wavefunction.

The rest of this note elaborates on some of these points. We shall first give a brief review of the debate, which originated (1) due to the puzzling property of the HH wavefunction applied to the quantum creation of the universe; and (2) from the discrepancy between the instanton approach and the naive WKB approach when applied to this problem. We then elaborate on some of the above arguments on why the instanton method is correct. We discuss the issue in terms of the Wheeler-DeWit (WDW) equation, where the cosmic scale factor $a$ plays the role of the coordinate, and re-examine the physics in the WDW equation.

\section{The Debate}

In ordinary quantum mechanics one can evaluate the wavefunction of
a system by evaluating a path integral with given initial conditions.
\baray
\psi(x,t) = <x,t|x_i, t_i> = N \int_{x_i,t_i}^{x, t} D[x] e^{iS[x,t]} 
\earay
The functional integral is between all possible world lines interpolating
 between the two boundary conditions. Upon Wick rotation one simply 
picks up the ground state contribution to the above propagator. In 
other words, $\psi(x,t)$ becomes the ground state wavefunction. One can
do the analogous in quantum gravity and define a ground state wavefunction
of the universe. The Hartle-Hawking proposal for the wavefunction of the 
universe imposes a certain initial condition in the Euclidean spacetime.
 The initial condition comprises of a null geometry and the final geometry 
is some three geometry, $h_{ij}$, and the interpolating geometries are compact
four geometries. 
\baray
\Psi(h_{ij}) = \int_{\emptyset}^{h_{ij}} D[g_{\mu,\nu}] e^{-S_E[g_{\mu, \nu}]} 
\earay 

(The conventions we follow are those of \cite{Hartle:1983ai}). The 
Euclidean action is defined as
\ba
\label{act}
S_{E} =-\frac{1}{16 \pi G}\int d^4 x \sqrt{-g} \left( R - 2\Lambda\right)
\ea
The Euclidean metric is given by
\ba
\label{mini}
ds^2 = d\tau^2 + a(\tau)^2 d\Omega_{3}^2 
\ea
With this metric ansatz, the 
action becomes
\ba
\label{action}
S_{E} = \frac{1}{2}\int d\tau \left( -a\dot{a}^2 - a + \frac{\Lambda}{3} 
a^3 \right)
\ea
The Hubble constant squared is given as usual by $H^2 = \Lambda / 3$.
The dominant contribution to the path integral comes from the saddle point
which can be found by looking for the solution of the Euclidean Einstein 
equation for this minisuperspace model. The result is the $S^4$ instanton
given by
\baray
\label{s4}
a(\tau) = \frac{1}{H} \cos (H \tau)
\earay

This instanton is interpreted as describing the creation of the universe
from a state of nothing. Pictorially this is represented in Fig. 1. The
lower half shows half the $S^4$ instanton (in Euclidean spacetime), 
the upper half shows a deSitter universe (in Minkowski spacetime). 

To find the probability of the formation of the deSitter universe
by this $S^4$ instanton one calculates the Euclidean action for
the instanton, i.e. $S_E[a(\tau)]$ where $a(\tau)$ is given by 
Eqn.(\ref{s4}). The probability is then simply given by the
\baray
\label{HH}
P_{HH} = e^{3\pi/G\Lambda} 
\earay

Assuming that a stringy landscape of vacua exists, one might
want to apply this wavefunction as a selection principle to choose
a particular deSitter vacuum. Then the above wave function seems to select
a vacuum with a vanishing cosmological constant. However, this is
precisely the value of $\Lambda$ where the wavefunction diverges.
This is an infrared divergence. \\

To get around this problem of the Hartle-Hawking proposal, Linde and
Vilenkin have proposed alternatives. There are two separate ways 
to motivate the Linde-Vilenkin wavefunction. 
In fact, technically (see \cite{Vilenkin:1998rp}) the tunneling 
wavefunction is distinguished from the Linde wavefunction. The reason 
we put them under the same classification is that they yield the same 
answer for the probability for the creation of the universe. \\

To introduce the LV wavefunction, let us consider a closed 
Robertson-Walker universe filled with a vacuum energy $\rho_{vac}$ 
. Let $a$ be the scale factor. The evolution equation for $a$ can 
be written as
\begin{align}
\pi^2 + a^2 - a^4/a_0^2 = \epsilon
\end{align}
where $\pi = -a\dot{a}$ is the momentum conjugate to $a$ and $a_0 = 
(3/4)\rho_{vac}^{-1/2}$. This equation is identical to that of a 
particle of energy $E = 0$ moving in a potential $U(a) = a^2 - 
a^4/a_0^2)$ (see Fig.$(2)$). 
In quantum cosmology, there is the possibility of 
the universe tunneling through the potential barrier 
to the regime of unbounded expansion. The semiclassical tunneling 
probability is estimated by Vilenkin using the WKB method
\begin{align}
\label{prob}
P \sim \exp \left( -2\int_{a_1}^{a_2}|\pi(a)|da \right ).
\end{align}
This gives the tunneling probability from nothing 
to a closed universe of a finite radius $a_0$. The corresponding 
probability can be found by setting $a_2 = a_0, a_1 = 0$ in Eq.(\ref{prob})
. The result is $P \sim \exp(-3\pi/G\Lambda)$, where $\Lambda$ and $
\rho_{vac}$ are related as usual. The tunneling approach to quantum 
gravity assumes that the universe originated in a tunneling event of 
this kind. Once it nucleates, the universe begins a deSitter expansion 
phase.\\

 The Wheeler-DeWitt equation for this simple model can be obtained 
by replacing $\pi(a) \to -id/da$. The resulting equation is
\begin{align}
\label{wheel}
\left(\frac{d^2}{da^2} -a^2 + \frac{a^4}{a_0^2} \right) \Psi(a) = 0
\end{align}
This equation has outgoing and incoming wave solutions corresponding to 
expanding and contracting universes in the classically allowed range 
$a > a_0$, and exponentially growing and decaying solutions in the 
classically forbidden range $0 <a < a_0$ (Fig.($2$)). 
The boundary condition that selects the tunneling wavefunction requires 
that $\Psi$ should include only an outgoing wave at $a \to \infty$. 
The under-barrier wavefunction is then a linear combination of the 
growing and decaying solutions. The two solutions have comparable 
magnitudes near the classical turning point $a = a_0$, but the 
decaying solution dominates in the rest of the under-barrier region.\\

\begin{figure}
\begin{center}
\epsfig{file=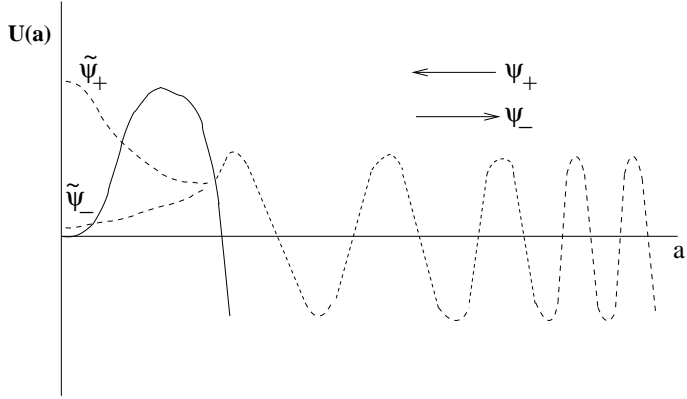, width=9cm}
\vspace{0.1in}
\caption{The tunneling wavefunction. The growing and the decaying 
components under the barrier combine to form the outgoing 
and incoming waves on the right.
If only the growing mode is present (the Hartle-Hawking wavefunction),
or if only the decaying mode is present (the Linde wavefunction),
there will be both outgoing and incoming waves on the right. 
Vilenkin demands a particular combination so there is only outgoing wave.
Decoherence will yield the outgoing wave as the classical solution.}
\label{fig2}
\end{center}
\end{figure}

In a more realistic model, the constant vacuum density $\rho_{vac}$ is 
replaced by the potential $V(\phi)$ of some scalar field $\phi$. If 
$V(\phi)$ is sufficiently slowly varying in $\phi$, one finds the 
same result as above, with the replacement $\rho_{vac} \to V(\phi)$.  
The highest probability is found for the largest values of $V(\phi)$ (
and smallest initial size $a_0$). Thus, the tunneling wavefunction 
predicts that the universe is most likely to nucleate with the largest 
possible vacuum energy. \\

  A similar proposal comes from Linde. Linde suggested that the 
wavefunction of the universe is given by a Euclidean path integral, 
with the Euclidean time rotation performed in the opposite sense, 
$t \to +i\tau$, yielding
\begin{align}
\Psi = \int^{(h_{ij},\phi)} e^{+S_E[h_{ij}, \phi]}
\end{align} 
where $h_{ij}$ is the metric of the $3$-geometry that we are interested 
in and the $3$-geometry contains some field $\phi$.
For the simple metric under consideration, this wave function gives 
the same nucleation probability as the tunneling wave function. \\

There is one big problem with this proposal. The path integral, 
since it is done with the opposite Wick rotation, is divergent 
over the matter and the inhomogeneous metric modes. This would 
lead to a catastrophic particle creation after the nucleation of 
the universe. \\

Although both the Vilenkin (tunneling) 
and the Linde proposals give the same answer for the simple model at 
hand, there is difference in them in terms of the boundary conditions 
they impose (see \cite{Vilenkin:1998rp}). As we have already mentioned, 
the Wheeler-DeWitt equation (\ref{wheel}) allows for two independent 
solutions (the growing and the decaying modes) under the barrier, and 
two independent solutions (the incoming and the outgoing modes) in the 
classically allowed region. Vilenkin's proposal includes a linear 
combination of the growing and the decaying modes under the barrier, 
and only an outgoing mode in the classically allowed region; whereas 
the Linde proposal includes only the decaying mode under the barrier 
and a linear combination of the incoming and the outgoing modes in 
the classically allowed region. Of course, for both the proposals, 
the decaying mode is dominant near $a = 0$. This is in keeping with 
the classical intuition that before tunneling the system (which, in 
this case, is the universe) should be localized at the origin. \\

Both the Linde and the Vilenkin proposals favor the nucleation of a 
universe with the largest value of $\Lambda$, which is in some sense is 
problematic in itself. The semiclassical description breaks down at values of 
$\Lambda$ close to the Planck scale. In fact, according to their proposal, 
quantum foam is more favored over a universe that is closed and inflating.

\section{Inclusion of a scalar field}

In any realistic model, the universe will contain both gravitational and 
non-gravitational degrees of freedom. Let us go beyond the minisuperspace
(that contains the scale factor $a(\tau)$ only) by including a scalar field
$\phi$., for example, the inflaton field. Consider Linde's proposal for the wavefunction which involves
Wick rotating the time axis in the opposite sense $t \to i\tau$. 
Such a rotation changes the sign in front of the Euclidean action in 
Eq.(\ref{act}), and the probability of the creation of the universe
is given by $P \sim e^{- 3\pi/G\Lambda}$. This proposal has the problem
that even though it cures the infrared divergence of the Hartle-Hawking
proposal, it makes the scalar field unbounded from below.\\

To see this, let us consider the scalar field action. In Loretzian 
signature the action is
\baray
S[\phi] = \int dt \left( \frac{1}{2}(\partial_{\mu}\phi)^2 -
 U(\phi)\right)
\earay 
We assume that the scalar field potential is bounded from below.
Now, under the usual Wick rotation $t \to -i\tau$, the path
integral is given by
\baray
\int D[\phi] e^{-S_E[\phi]}
\earay
where $S_E[\phi]$ is given by
\baray
S_E[\phi] = \int d\tau  \left( \frac{1}{2}(\partial_{\mu}\phi)^2 +
 U(\phi)\right)
\earay
Since, $U(\phi)$ is bounded below and in Euclidean signature 
the kinetic term is positive definite, the Euclidean action is
also bounded from below. However, if one does the opposite
Wick rotation $t \to i\tau$, then the action ceases to be bounded
from below after the Wick rotation. The path integral is given by
\baray
\int D[\phi] e^{|S_E[\phi]|}
\earay
Consequently, the scalar field becomes unstable. This was
noted in \cite{Lapchinsky:1979fd}. The scalar field becomes
unstable and leads to a catastrophic particle production.

\section{The Modified Wavefunction}

One of the main motivations for looking for alternatives (i.e., solutions to
the WDW equation with different boundary conditions) to the
Hartle-Hawking wavefunction as a possible description of the spontaneous
creation of the universe was its preference for a universe with a 
vanishing $\Lambda$. In this section we shall argue that once 
one goes beyond the minisuperspace model, it is natural to expect
corrections that will cure this infrared divergence of the wavefunction.

In \cite{Halliwell:1985eu} the authors calculated the metric and scalar
field perturbations. The homogeneous degrees of freedom corresponding to 
the scale factor and the scalar field were treated exactly. The 
inhomogeneous degrees of freedom were treated to the second order.
These inhomogeneities were shown to lead to the temperature anisotropies
of the order observed in the CMB. If we sought a description solely in terms
of the homogeneous degrees of freedom, then these inhomogeneities were
shown to lead to a redefinition of the cosmological constant, the
Newton's constant, and the generation of curvature squared counter terms.
This is just renormalization of the parameters in the action such that
the UV cut-off dependent physics does not show up in the low energy
description.\\

We shall argue in this section that the above is true if one constrains
the symmetry of the $S^4$ instanton (i.e., the deSitter $SO(5)$ rotational 
symmetry) to remain unbroken upon the inclusion of the inhomogeneous 
degrees of freedom. However, there is no apriori reason for this. 
In fact, for a pure deSitter spacetime, where the cosmological constant
comes from a scalar field with a flat potential, the large wavelengths
of the scalar field fourier modes show up as deSitter fluctuations and couple
with the scale factor.  We shall
show that this backreaction leads to natural metric fluctuations 
in a deSitter space which 
lead to a breaking of the $SO(5)$ symmetry and the generation of a radiation
term on the inclusion of the inhomogeneous modes (these inhomogeneous
modes also lead to a redefinition of the curvature constant $K$ present
in the Friedmann equations. But , as usual, $K$ can be reset to $+1$ for
the closed universe). In the absence of these
metric perturbations the zero point energies of the inhomogeneous modes
leads only to the renormalization of the cosmological constant, the Newton's
constant, and the curvature squared counterterms. But the metric perturbations,
 natural in a deSitter spacetime, lead to an additional radiation term.
These metric perturbations are similar to the ones discussed in, for example,
\cite{Starobinsky:1986fx} in the context of stochastic inflation. There
the regime where one obtains a Hartle-Hawking like distribution for
the scalar field is also the regime where the metric has sizeable
fluctuations.
\\

 As discussed in Sec. $1$, the Euclidean action for the minisuperspace model
(for which the metric is given by Eq.(\ref{mini})) is given in terms of the
scale factor by Eq.(\ref{action}). This, in turn, leads to the Hartle-Hawking
result Eq.(\ref{HH}). Let us include metric perturbations. The perturbed
three metric is given by
\baray
\label{pert}
h_{ij} = a^2 \left( \Omega_{ij}+\epsilon_{ij}\right)
\earay
where $\Omega_{ij}$ is the metric on the unit three-sphere and 
$\epsilon_{ij}$ is a perturbation on this metric and can be expanded
in spherical harmonics over $S^3$ \cite{Halliwell:1985eu}. For simplicity
let us consider only the tensor perturbations.
The tensor perturbations, with amplitudes $t_n$, have the Euclidean action
\ba
S_E^n = \frac{1}{2}\int t_n \hat{D} t_n +  ~boundary ~term
\ea
where
\ba
\hat{D} = \left(- \frac{d}{d\tau}\left[a^3\frac{d}{d\tau} \right]
+ a(n^2 - 1)  \right)
\ea
where the background satisfied the classical equation of motion.
The action is extremized when $t_n$ satisfies the equation
$\hat{D} t_n = 0$, i.e. the equation
\ba
\label{eom_ten}
\frac{d}{dt}\left[a^3 {\dot{t_n}} + (n^2 -1)a t_n\right] = 0
\ea
For $t_n$ that satisfies the equation of motion, the action
is just the boundary term
\ba
\label{E_cl}
S_E^{n(cl)} = \frac{1}{2}a^3 \left(t_n\dot{t_n}+4\frac{\dot{a}}{a}
t_n^2  \right)
\ea 
The path integral over $t_n$ will be
\ba
\label{tensor}
z_n = \int d[t_n]\exp (-S_E^{n(cl)})= (\det \hat{D})^{-1/2}\exp (-S_E^{n(cl)})
\ea
Now let us perform the path integral over the superspace comprising
the scale factor and the tensor modes. We shall get the following path
integral:
\baray
Z = \int D[a] e^{-S_E} \prod_n z_n
\earay
where $S_E$ is the Euclidean action for the scale factor given by
Eq.(\ref{action}). If we are interested only in the dynamics of 
the scale factor and not the high frequency tensor modes, then
we simply trace them out. Notice that the tensor mode classical action 
$S_E^{n(cl)}$ depends on boundary conditions. Let us label the initial
and the final value of the tensor perturbations as $t_n^i$ and $t_n^f$, 
respectively. To take the trace we set $t_n^i = t_n^f$, and integrate 
over all possible values of $t_n^i$. This is nothing but an integral over
the propagator with periodic boundary conditions and it gives the
sum over the various eigenvalues of the simple harmonic oscillator 
corresponding to the tensor mode (of course, with a variable mass and
frequency) \cite{colemanL, Felsager:1981iy}.
\baray
\int dt_n^i \int dt_n^f \delta (t_n^i - t_n^f) z_n = \frac{1}{e^{D_n/2} - e^{
-D_n/2}}
\earay
where $D_n$ is the integrated frequency and $D_n/2$ is the ground state 
energy of the oscillator,
\baray
\label{Dndef}
D_n = \int d\tau \frac{\sqrt{n^2 -1}}{a(\tau)} \simeq  
\int d\tau \frac{n}{a(\tau)}
\earay
The denominator $a(\tau)$ in the frequency is simply the redshift factor.
So tracing out the tensor perturbations, one gets the path integral
\baray
\int  D[a] e^{-S_E}\prod_n \frac{1}{e^{D_n/2} - e^{-D_n/2}}
&\simeq& \int  D[a] e^{-S_E}\prod_n \left( e^{-D_n/2}+  e^{-3D_n/2} + ... \right) \nonumber \\
&\simeq& \int  D[a] e^{-S_E - \sum_n D_n/2} = \int  D[a] e^{-S_E^{mod}}
\label{act1}
\earay
where only the dominant term is kept.

Let us now look at the counting of modes.
From Eq.(\ref{Dndef}) and Eq.(\ref{act1}), it is clear that the modified action is given by
\baray
\label{act2}
S_E^{mod} = S_E + \sum_n D_n/2 = \frac{1}{2}\int d\tau \left( 
-a\dot{a}^2 - a + \frac{\Lambda}{3} a^3  + \frac{1}{a(\tau)}\sum_n n \right)
\earay
The index $n$ really stands for ${n,l,m}$ that represent the spherical
harmonics on $S^3$. As explained in \cite{Sarangi:2005cs}, this leads 
to the following result
\baray
\label{sum}
\sum_n n \simeq \int_0^{N}dn n ~ \times~ n^2 = \frac{N^4}{4}
\earay
where the factor of $n^2$ shows the degeneracy arising due to the
indices $l$ and $m$ for a given $n$. $N$ is given by
\baray
\label{N}
N = \frac{H^{-1}}{l_s} f(a)
\earay
that is, we include the wavelengths that lie between string scale 
(UV cut-off), and the Hubble scale (IR cut-off). The function $f(a)$
is a fourth degree polynomial in $a(\tau)$ that we shall discuss now.\\

\begin{itemize}
\item Maintaining the $O(5)$ symmetry : \\
The simplest counting is given by $f(a) = a$ in Eq.(\ref{N}). The radius
of $S^3$ grows from zero (at the south pole of $S^4$) to a maximum value
at the equator, to back to zero at the north pole. So when one counts the
modes, one might simply take $N = \frac{H^{-1}}{l_s} a(\tau)$.

From Eq.(\ref{act2}) one can see that this leads to a renormalization of 
the cosmological constant from the ground state energies of the tensor modes:
\baray
\Lambda \to \Lambda + \frac{N^4}{4} 
\earay 
where $\frac{N^4}{4} \equiv \frac{1}{4(H l_s)^4}$.
Of course, a proper calculation using renormalization techniques will also
give $\log (H/M_P)$ terms. This is well known.

\item Breaking the $O(5)$ symmetry: \\
However there is no reason to restrict oneself to maintaining the
$O(5)$ symmetry. In fact, as explained in \cite{Halliwell:1985eu},
to mainitain the $O(5)$ symmetry, the perturbations must vanish at the
poles of the $S^4$ geometry. Since we are considering only the tensor
modes, $t_n = 0$ at the poles. This is very much possible at the classical
level. However, once we treat the perturbations quantum mechanically,
 then even assuming that the perturbations start off in their ground states
at the south pole of $S^4$ (as in \cite{Halliwell:1985eu} ) it is easily
seen that the amplitudes of the perturbations do not vanish at the poles.
The amplitude of the oscillators has a nonzero value in their ground states.
In fact, as one approaches the poles, the oscillator conributions backreact
the geometry more and more till the WKB approximation used in 
\cite{Halliwell:1985eu} breaks down. One expects some kind of flattening
of the poles due to this backreaction. This will lead to the breaking
of the $O(5)$ symmetry.\\

If we allow for this breaking of the symmetry, then $f(a)$ will assume
a general form:
\baray
N^4 \propto f(a)^4 = A a^4 + B a^3 + C a^2 + D a + E
\earay
 One can see from Eqs.
(\ref{act2}, \ref{sum}) that the $E$-term will lead to a radiation
term in the modified action. The $Da$ term will lead to a matter term
. Such a matter term was anticipated in \cite{Firouzjahi:2004mx} and leads
to a $\frac{1}{\Lambda^{3/2}}$ correction to the Hartle-Hawking result.
The $C a^2$ term leads to a redefinition of the curvature. Recollect
that if we retain a curvature $K$ term from the Friedmann equation, then
the $a$ term in Eq.(\ref{act2}) should really be written as $Ka$. That is,
the unmodified Lagrangian is $\frac{1}{2}(-a\dot{a}^2 - Ka + \frac{\Lambda}
{3} a^3)$. 
The $C a^2$ term just leads to a redefinition of $K$, which can be 
reset to $K = + 1$ for closed universe such as the one we are studying.
The $A a^4$ term, as we saw in the previous discussion, simply 
renormalizes the cosmological constant.  From the previous discussion we 
expect $A = 1$. As we shall explain now, we expect the $B$ term
to be equal to zero.  \\

To get an estimate of these coefficients, let us consider a more familiar
situation in an inflationary universe where the cosmological constant is 
supplied by a scalar field potential
$V(\phi)$. During the deSitter phase, the kinetic term is negligible.
However, we know that in a deSitter spacetime there are fluctuations
of scalar field given by $\delta \phi = H/2\pi$. Unless the inflaton
field is slowly rolling close to the end of the inflation, the 
inflaton will typically undergo fluctuations of the order $H$ in a time
 period $H^{-1}$ \cite{Linde:2005ht}. So $\dot{\phi} \sim H^2$. 
The corresponding metric fluctuation is given by 
$\delta a/a = H \delta \phi / \dot{\phi} \simeq 1$. This can be seen
in terms of metric perturbations. Let us choose a gauge in which
the metric perturbations occur only in the $g_{ij}$ - i.e.
the spatial components. So the unperturbed metric goes to 
a perturbed one
\baray
a^2 \delta_{ij}dx^idx^j \to a^2 [(1+2D)\delta_{ij} + 2E_{ij}]dx^idx^j
\earay
where we are following the notations and treatment of \cite{Liddle:2000cg}.
One can separate the above perturbations into scalar and tensor
parts as follows,
\baray
a^2 [(1&+&2D)\delta_{ij} + 2E_{ij}]dx^idx^j  \\
&=& a^2\delta_{ij}dx^idx^j + 2Da^2\delta_{ij}dx^idx^j + 2a^2E_{ij}dx^idx^j
\nonumber \\
&=& a^2dx^idx_i +2Da^2\delta_{ij}dx^idx^j + 2a^2 \left(-\frac{k_ik_j}{k^2} 
+ \frac{1}{3}\delta_{ij} \right)E dx^i dx^j \nonumber \\
&=& a^2 dx_idx^i + 2a^2 \left(D +\frac{1}{3}E \right) \delta_{ij}dx^idx^j - 2Ea^2\frac{k_ik_j}
{k^2}dx^idx^j \nonumber 
\earay
The scalar metric perturbation $R$ is given by
\baray
R = D +\frac{1}{3}E
\earay
Considering only the scalar perturbation to the scale factor, we can write
\baray
R = \delta(a^2)/2a^2 = \delta a /a
\earay
Furthermore, the term $D +\frac{1}{3}E$ is related to the
scalar field perturbation by
\baray
D +\frac{1}{3}E = H\frac{\delta \phi}{\dot{\phi}}
\earay
Hence
\baray
\delta a/a = H \delta \phi / \dot{\phi} \simeq 1
\earay
So, $\delta a \simeq 
H^{-1}$ over one Hubble time. To get an estimate
of the coefficient $E$, let us write $f(a)^4$ as
\baray
f(a)^4~ =~ <(a(\tau) + \delta a)^4> 
\earay 
where $<...>$ is the expectation value. Since the de Sitter fluctuations
are random, we have $<\delta a> = <\delta a^3> = 0$. So 
we simply have a radiation term apart from the renormalization of
$\Lambda$ and the curvature term $K$. Here, $E \sim 1$. 

The metric perturbation may be studied in the same way. This tensor mode
fluctuation has amplitude $v_{k}$. In an inflationary universe, such modes
leave the horizon as $k=aH$, which is then frozen until they re-enter the horizon.
They may be detected via their effect on the polarization of the cosmic 
microwave background radiation. Since the tensor mode is responsible for the 
primordial B mode polarization, detecting the B mode is the main goal of future
CMB observations. In our calculation, we are simply integrating over this tensor mode
fluctuations. These tensor modes are massless, so they contribute to the radiation in the deSitter 
space, thus breaking the deSitter symmetry. This is spontaneous symmetry breaking 
by radiative corrections. In Euclidean metric, they modify the $S^{4}$ instanton in a way 
such that the $SO(5)$ rotational symmetry is broken.
Since the universe emerging from the tunneling is a closed universe, there are no modes with wavelength longer than the size of the universe, i.e., $\Lambda^{-1/2}$. So there are no superhorizon modes. 
The divergence at $k \rightarrow \infty$ requires the introduction of 
a cut-off $M_{s}$, which is naturally provided by the string scale in string theory. 
So the sum over modes yields a term $\sim M_{s}^{4}/\Lambda^{2}$.

Since a pure quantum deSitter space does not exist, we cannot consider its tunneling from nothing.
Instead, consistency requires us to consider the tunneling to a deSitter space with a radiation term.
This radiation term receives contributions from the metric perturbation as well as from the quantum fluctuations of all matter fields, so it depends on the specific vacuum the tunneling process is heading. 
This leads to the modified action
\baray
\label{act3}
S_E^{mod} \equiv - F =  \frac{1}{2}\int d\tau \left( 
-a\dot{a}^2 - a + \frac{\Lambda}{3} a^3  + \frac{9 b M_s^4}{4a(\tau) 
\Lambda^2}\right)
\earay 
where $M_s$ is the string mass scale, and $\Lambda$ is now the redefined 
cosmological constant. Here $b$ measures the degrees of freedom in the 
approximate deSitter space. The
curvature $K$ has been set to $+1$. We have defined $F$ here as 
a ``free energy''. The inclusion of the radiation term generalizes 
the entropy to free energy (free energy per unit temperature, to be
precise). The equation of motion is now given by the variation of the
modified action (setting $b=1$ for the moment) \cite{Sarangi:2005cs}
\ba
\label{modbounce}
-2\frac{\ddot{a}}{a} - \frac{\dot{a}^2}{a^2}+\frac{1}{a^2}
= \Lambda - \frac{9M_s^4}{4\Lambda^2 a^4} 
\ea
the solution to which is given by
\ba
\label{bouncesol}
a(\tau) = \frac{1}{\sqrt{2}H} \sqrt{\left( 1 
+ \sqrt{\left( 1 - \frac{2GM_s^4}{3\pi H^2}\right)} \cos 
(2 H\tau) \right)}
\ea
This is an interesting geometry. Although it has radiation, it does not
encounter any singularity.

\begin{figure}
\begin{center}
\epsfig{file=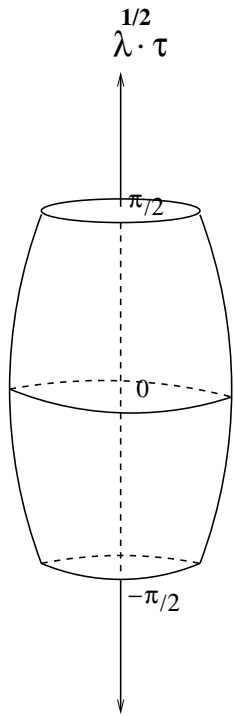, width=3cm}
\vspace{0.1in}
\caption{The Modified Bounce : It is a squashed version of $S^4$ resulting in a ``barrel''. 
Topologically, it is still a $S^{4}$. Presumably, higher order corrections will smooth the edges 
of the barrel, probably resulting in an ellipsoid.}
\label{fig6}
\end{center}
\end{figure}

The modified result is then given by
\baray
P \simeq e^{-F} = \exp \left(\frac{3\pi}{G\Lambda} - \frac{6M_s^4}{\Lambda^2} \right)
\earay
The probability of the nucleation of the universe from nothing is now
determined by {\it the minimization of a free energy}.
\end{itemize}


\section{A Problem of the Hartle-Hawking Wavefunction and Its Possible Resolution}

Recall that the HH wavefunction for deSitter space is given by 
\baray
\label{HH}
\Psi_{HH} \simeq \exp ({3\pi/2G\Lambda}) 
\earay
For dynamical $\Lambda$, this implies that $\Lambda \rightarrow 0$ is much preferred.
Since the cosmic scale factor $a$ which measures the size of the universe, behaves as 
$a \simeq 1/\sqrt{\Lambda}$, this implies the creation of an infinite size (Minkowski) universe 
is exponentially preferred. 
Based on our knowledge of quantum mechanics, a macroscopic object behaves classically, so tunneling, a quantum process, should be very much suppressed. 
So this poses a puzzle. Phenomenologically, our universe has gone through the big bang epoch and most probably an inflationary phase before that, so the prediction of the HH wavefunction is 
inconsistent with observation. 

The above intuitive argument has been laid out in many papers \cite{Coleman:1988tj}.  Clearly, there is a serious problem associated with the HH wavefunction. Here, we like to point out that this problem associated with the HH wavefunction may disappear when it is replaced by the improved wavefunction.
Let us first review the problem and then show how it may be resolved.

To quantify this problem more precisely, let us consider the argument in Ref.\cite{Fischler:1990se}.
If tunneling via a $S^{4}$ instanton has such a large probability, then tunneling via a 
chain of connected $S^{4}$s is even more likely, 
with wavefunction
\baray
\Psi  \sim \exp \left(k \frac{3\pi}{2G\Lambda \hbar} - [k-1] \frac{\Delta}{G \hbar} \right)
\earay
where $\Delta$ is the contribution from the small neck connecting the $S^{4}$s, $\Delta \sim r^{2}$, with $r$ the size of the neck, and $k$ is the number of $S^{4}$s in the chain. Any reasonable estimate of $\Delta$ yields a finite value that depends only weakly on $\Lambda$. Although $\Delta$ leads to some suppression, the first term in the exponent dominates over the $\Delta$ term as $\Lambda \rightarrow 0$. 
As a result, the tunneling probability $|\Psi|^{2}$ is dominated by the large $k$ contributions. This leads to a pathology as $k \rightarrow \infty$. 

Even if $k$ has a maximum (one may argue that $k_{max} \sim \exp (3\pi/G\Lambda)$), the HH wavefunction has a serious problem. Consider a point $p$ in one of the deSitter spaces and compare the amplitude for the geomtery at $p$ to be smooth versus that to find a Schwarzschild wormhole end ( i.e., a black hole)  at $p$. The leading term overwhelms the suppression so the result predicts that black holes are dense in Minkowski spacetime. As pointed out in Ref.\cite{Fischler:1990se}, this is the best case scenario. It could be that the wavefunction is too divergent to make any sense. 

The above disaster probably happens to any quantum theory where the $S_{E}$ is unbounded from below.
It is usually assumed that the most symmetric instanton solution yields the lowest Euclidean action.
This is clearly violated here. This is related to the unboundedness from below of the $S^{4}$ Euclidean action.

 Now consider the improved wavefunction. Here, we now have
\baray
\Psi  \simeq \exp \left(-k F - [k-1] \frac{\Delta}{G \hbar} \right)
\earay
Now, if $F+ \Delta / G > 0$, or 
\ba
\label{inequal}
\Delta > -GF =\frac{3\pi}{2\Lambda} - \frac{6M_s^4G \hbar }{2\Lambda^2}
\ea
then the single barrel term, that is, $k=1$ term dominates. If $\Delta$ is independent of $\Lambda$, then
the right hand side of the above inequality can be replaced by its value at $\Lambda_{max}$, the value  that maximizes the tunneling rate for a single barrel, i.e., $\Delta > {3\pi}/{4\Lambda_{max}}$.
If $\Delta$ depends on $\Lambda$, the above inequality must be 
satisfied for all allowed $\Lambda$, not just $\Lambda_{max}$. That is, if the inequality (\ref{inequal}) is not satisfied for some allowed $\Lambda$, then the tunneling probability will increase as $k$ increases, so for $k \rightarrow \infty$, the tunneling rate blows up.
 
It is difficult to estimate $\Delta$ reliably. To get an idea of the dimensionality of $\Delta$ here, suppose $r \simeq a$ at the end of the barrel. This yields $\Delta \sim a^{2} \sim 1/(GM_{s}^{2})^{2}$. Since $F$ scales in the same way in general, $F \simeq 1/(GM_{s}^{2})^{2}$, a more careful analysis is required to see if $F+ \Delta / G > 0$ is satisfied or not. 
So we see that the improved wavefunction may not suffer from the problem that plagues the HH wavefunction. Clearly, a more careful examination of this issue will be very important.

In usual field theory, the tunneling probability is dominated by the most symmetric instanton.
If this is the case here, it would have been the $S^{4}$ instanton. Instead, as argued in Ref\cite{Coleman:1988tj}, this is not the case here. This leads the consideration of wormholes and the above chain of bubbles. This is a consequence of the unboundedness of the Euclidean action of the $S^{4}$ instantoin. In contrast, the barrel action has a lower bound, and so offers a chance to avoid the disaster encountered by the $S^{4}$ instanton. In contrast, the single barrel is in some sense more symmetric than the chain of bubbles. This offers the hope that the single barrel dominates the tunneling probability.

\section{Hartle-Hawking Distribution from Stochastic Inflation}

 We shall give a quick review of stochastic inflation. The reader should
refer to \cite{Starobinsky:1986fx,Linde:2005ht,Linde:1993xx} for
 details. The basic idea is that the dynamics of a large-scale 
quasi-homogeneous scalar field producing the de Sitter stage is 
strongly affected by small-scale quantum fluctuations of the same scalar 
field and, hence, becomes stochastic. The evolution of the 
corresponding large-scale space-time metric follows that of the 
scalar field and is stochastic too. One can write down a Langevin
equation for the long wavelength modes of the scalar field and a
corresponding Fokker-Planck equation for the  distribution function
for the large wavelength modes.

Let the de Sitter phase be produced by the potential of some
scalar field with the Lagrangian density
\baray
L = \frac{1}{2}\partial_{\mu} \phi \partial^{\mu}\phi - V(\phi)
\earay 
The Hubble constant is given by
\baray
H^2 = \frac{8\pi G V(\phi)}{3}
\earay
At the de Sitter phase $H \simeq H_0 =$ constant, and the scale factor
grows as $a(t) = a_0 \exp (H_0 t)$. It is natural during the de Sitter
phase to separate the full scalar field $\phi$ into a long wavelength
part $\bar{\phi}$ and the short wavelength part 
\baray
\label{phi}
\phi = \bar{\phi}(t, \vec{r}) + \int \frac{d^3k}{(2\pi)^{3/2}} ~
\theta(k - a(t)H_0)~ [\hat{a}_k \varphi_k(t) e^{-i\vec{k} \cdot \vec{r}}
+ \hat{a}_k^{\dagger} \varphi_k^*(t) e^{i\vec{k} \cdot\vec{r}}] 
\earay
where $\hat{a}_k^{\dagger}$ and $\hat{a}_k$ are the usual creation and 
annihilation operators.
Here $\bar{\phi}(t, \vec{r})$ contains only the long wavelength modes
with $k < H_0 a(t)$. The second integral term satisfies the free massless
scalar wave equation in the deSitter background : $\Box \varphi = 0$
(assuming that the mass of the scalar field is much smaller than $H$). 
The solution is well known 
\baray
\label{varphi}
\varphi_k = H_0 (2k)^{-1/2}\left( \eta - \frac{i}{k} \right)\exp(-ik\eta);
\eta = \int\frac{dt}{a(t)} = -\frac{1}{a(t)H_0}
\earay
 The scalar field $\phi$ satisfies the operator equation of motion
$\Box \phi + \frac{dV}{d\phi} = 0$ exactly.  Using Eq.(\ref{phi},\ref{varphi}) 
and the slow roll conditions, one obtains the following
equation of motion for $\bar{\phi}$ in the leading order
\baray
\label{barphi}
\dot{\bar{\phi}}(t,\vec{r}) = -\frac{1}{3H_0}\frac{dV(\bar{\phi})}{d\bar{\phi}}
+ f(t, \vec{r})
\earay
where $f(t, \vec{r})$ is a ``noise'' term given by
\baray
f(t,\vec{r}) = -i\frac{a(t)H_0^2}{(2\pi)^{3/2}} \int d^3k~ \delta(k - aH_0)
~ \frac{H_0}{\sqrt{2}k^{3/2}} [\hat{a}_k \exp(-i\vec{k}.\vec{r})
- \hat{a}_k^{\dagger} \exp(i\vec{k}.\vec{r})] 
\earay
The large scale scalar field $\bar{\phi}$, therefore, changes not only
due to the classical force $dV(\bar{\phi})/d\bar{\phi}$ but also due 
to the flow of initially small-scale quantum fluctuations across the 
deSitter horizon $k = a(t) H_0$ in the process of expansion. This, then,
is a stochastic process.  Eq.(\ref{barphi}) is the Langevin equation 
governing the evolution of $\bar{\phi}$. 
The evolution of inhomogeneous modes is linear
inside the deSitter horizon ;  on the other hand, the evolution of 
$\bar{\phi}$ is non-linear. Note that there are no spatial derivatives
in Eq.(\ref{barphi}) at all. This means that if we are just interested 
in the evolution of $\bar{\phi}$, and not in what goes on inside the
Hubble patch, then we can consider the Hubble patch as a point.  
The temporal evolution of $\bar{\phi}$ is slow compared to $H^{-1}$
 so only processes with characteristic time $H^{-1}$ need to be
considered. 

Even though $\bar{\phi}$ and $ f(t,\vec{r})$ have a complicated 
operator structure, all terms commute with each other, so we can consider
$\bar{\phi}$ and $f(t, \vec{r})$ as classical. But they are stochastic
because one cannot assign any definite numerical value to the combination
$[\hat{a}_k \exp(-i\vec{k}.\vec{r}) - \hat{a}_k^{\dagger} \exp(i\vec{k}.
\vec{r})] $. The two point function for the noise term can be calculated
\baray
<f(t_1, \vec{r}) f(t_2, \vec{r})> = \frac{H_0^3}{4\pi^2} \delta(t_1 -t_2) 
\earay
Thus $f(t)$ has the properties of white noise.

One can also derive the Fokker-Planck equation corresponding to the 
Langevin equation for $\bar{\phi}$. One is interested in the average
value $<F(\bar{\phi})>$ of some arbitrary function $F$. One introduces
a normalized probability distribution $\rho(\bar{\phi}, t)$ , $\int_
{-\infty}^{\infty} \rho(\bar{\phi}, t) d\bar{\phi} = 1$, such that
\baray
<F(\bar{\phi})> = \int_{-\infty}^{\infty}\rho(\bar{\phi},t)F(\bar{\phi})
d\bar{\phi}
\earay
The Fokker-Planck equation for $\rho$  is given by
\baray
\label{fp}
\frac{d\rho}{dt} = \frac{H_0^3}{8\pi^2}\frac{\partial^2\rho}{\partial
\bar{\phi}^2} + \frac{1}{3H_0} \frac{\partial}{\partial \bar{\phi}}
\left( \frac{dV}{d\bar{\phi}} \rho\right)
\earay
One has to specify the initial condition $\rho_0$ at some initial time $t_0$
in order to solve this equation. One should note that this Fokker-Planck
equation is applicable only during the slow-roll stage. Later the second
time derivative will become important.\\

\subsection{The Hartle-Hawking Distribution from Stochastic Inflation}

One has to specify the initial condition $\rho_0$ in order to be able
to solve the Fokker-Planck equation for the distribution function
$\rho(\bar{\phi}, t)$. So how can one find something like the Hartle-
Hawking wavefunction which corresponds to ``no boundary''? A natural
idea is to look for stationary solutions  that are independent of
$t$. It is easy to solve Eq.(\ref{fp}) with $ d\rho/dt = 0$. One gets
the following solution
\baray
\rho = \rho_0 V(\phi)^{-1} e^{3\pi/G\Lambda}
\earay
where $\Lambda = 8\pi G V(\phi)$ is the cosmological constant.
This is precisely the square of the Hartle-Hawking wavefunction.

It is very remarkable that the stationary distribution coincides
with the distribution from Hartle-Hawking wavefunction. Eventually
the scalar field will start rolling in some Hubble patches and 
inflation will end in those patches. However, one should note that the
stationary distribution corresponds to that phase  when the noise
term $f(t,\vec{r})$  is of the same strength as the classical term
$dV/d\bar{\phi}$. This leads to the important conclusion that
the phase to which the HH distribution is applicable
is the one when the stochastic term is non-negligible. During
this phase the fluctuations in the metric is of the order of the
Hubble parameter $\delta a \sim H^{-1}$.  

\subsection{Metric Fluctuations during stochastic inflation and mode
counting}

The fact that the regime of stochastic inflation where the Hartle-
Hawking distribution is a solution of the Fokker-Planck equation 
also has big metric fluctuations will have important consequence on 
the mode counting. As was already noted in \cite{Starobinsky:1986fx},
the space-time metric during stochastic inflation can be written as
\baray
ds^2 = -dt^2 + e^{h(\vec{r})}a^2(t) (h_{ij}dx^i dx^j)
\earay
where $h(\vec{r})$ is not small and $a(t)$ is the scale factor for a 
strictly isotropic and homogeneous solution. The quantity $h(\vec{r})$
is essentially stochastic, its r.m.s. value being of the order of its
average. If the conditions of the slow-roll inflation are satisfied,
some Hubble patch eventually exits the stochastic inflation phase and rolls
down to end the inflation (this just the idea of eternal inflation).
During this last ``useful'' part of inflation, sufficiently large regions
are produced with the degree of perturbations that matches the observations.
During this part, the stochastic noise term in Eq.(\ref{barphi}) is negligible
and the evolution is governed by the classical force term $dV/d\bar{\phi}$.
This is just the slow-roll inflationary stage. The inhomogeneities 
produced at this stage are small and are given by
\baray
h(\vec{r}) = -2H_0
\delta \bar{\phi} / \dot{\bar{\phi}}
\earay
which gives the usual result $\delta a^2/a^2 = -2H_0 \delta \bar{\phi} / 
\dot{\bar{\phi}}$.

Hence, during the phase of stochastic inflation when the distribution
is given by the Hartle-Hawking result, the metric perturbations are
given by $\delta a / a \sim 1$. Whereas, it is only much later, during
the slow rolling phase, that the much smaller density perturbations
corresponding to observations are produced. 

This has important consequence for the mode counting. The number
of modes between the string scale and the Hubble scale were given by
\baray
N^4 = \left( \frac{H^{-1} a(t)}{l_s} \right)^4
\earay
This just led to the redefinition of the cosmological constant.
However, taking into account the metric fluctuations, the
mode counting is now done as
\baray
N^4 =  \left( \frac{H^{-4} <(a(t) + \delta a )^4> }{l_s^4} \right)
\earay

If the fluctuations are random, then $<\delta a>= <\delta a^3>=0$. 
This is the proper prescription for
mode counting for a purely stochastic fluctuation as this.
The result is a radiation term. This radiation term backreacts and 
modifies the $S^4$ instanton and gives a barrel solution.

\section{The Wheeler-DeWitt Equation and the Hartle-Hawking Wavefunction}

The Wheeler-DeWitt (WDW) equation is the Schrodinger like equation for 
the universe. Just like the Schrodinger equation is a way to impose
the Hamiltonian energy condition at the quantum level, the WDW equation
is a prescription to impose the Hamiltonian constraint at the quantum
gravitational level. 

The Wheeler-DeWitt (WDW) equation implements the Hamiltonian constraint at the
quantum level. It is given by
\baray
\label{wdw}
H \Psi = 0
\earay
where $H$ is the total Hamiltonian of the universe. 

The Hartle-Hawking
wavefunction is a solution of the WDW equation with the following boundary 
condition at $a = 0$
\baray
\label{bchh}
\Psi_{HH}(a=0) = 0
\earay
(See Fig.($2$)). In the classically allowed region Hartle-Hawking proposal
demands the presence of both the incoming wave and the outgoing wave.\\

The boundary condition Eq.(\ref{bchh}) has a few virtues to begin with.
First, the singularity at $a = 0$ is avoided by this boundary condition
\cite{DeWitt:1967yk,Hartle:1983ai}. Whereas, in the Linde and Vilenkin cases,
the wavefunction has a finite value at $a = 0$, and it is hard to see how
the universe would be metastable at $a=0$. In fact, when one considers the
quantum tunneling of a system from a metastable vacuum to a true vacuum
in usual quantum field theory, both the metastable vacuum and the true 
vacuum should be obtainable as solutions of the classical equation of motion. 
In such a case it would seem natural to impose a Linde/Vilenkin like
boundary condition at $a = 0$. 
As we prepare the system in a metastable vacuum state and leave it there
to sit classically (hence, the wavefunction would be nonvanishing 
at the metastable vacuum), we would like to ask for the probability of quantum
tunneling. However, for a deSitter universe (described by $a(t) = H^{-1} 
\cosh (Ht)$), $a = 0$ is not a classical solution 
for any value of $\Lambda = 3H^2$  and time. 
So it is hard to see the meaning of
a boundary condition where the wavefunction $\Psi(a=0)$ does not vanish.
Furthermore, a non-vanishing $\Psi(a=0)$ disagrees with our notion of a unique ``nothing''.
The Hartle-Hawking proposal (Eq.(\ref{bchh})) is more natural in this
situation. With $\Psi_{HH}(a=0) = 0$, we have a unique ``nothing''.


The landscape is a vast collection of vacua with different properties (like
the cosmological constant, various gauge field couplings, etc) \cite{Kachru:
2003aw,Bousso:2000xa,Douglas:2003um}.
In the context of the application of the wavefunction of the universe as
a selection criterion on the stringy landscape, Eq.(\ref{bchh}) is probably the only boundary condition that makes sense. In such a case, Eq.(\ref{bchh}) allows for a democratic 
comparison of various vacua. The no boundary proposal obviates the necessity
of any initial conditions \cite{Hartle:1983ai}. One does not have to
pick an initial vacuum in the landscape and then wonder why the 
universe ended up in that particular vacuum. 

\section{Some Comments}

There has been some recent works in applying quantum cosmology to the cosmic landscape
\cite{Kane:2004ct,Brustein:2005yn,McInnes:2005su,McInnes:2005ep,Huang:2005wq,
Davidson:1999fb, Holman:2005eu,Hawking:2006ur}. In \cite{Brustein:2005yn} 
the authors motivate the modified Hartle-Hawking wavefunction using
string thermodynamic arguments and the applicability of effective field theory
description. It is interesting to note that their result derived from rather general
arguments. It is possible that a stringy thermodynamic phase in the very
early universe might lead to a radiation like the component that we have
seen will modify the wavefunction. Ref\cite{Huang:2005wq} discusses
possible observational effects of the modified wavefunction and also 
that chaotic eternal inflation might be highly constrained by the use of 
the modified wavefunction. It is natural to believe that the usual initial state
of a scalar field in the Bunch-Davies vacuum will recieve corrections
due to the modified wavefunction. This will be akin to the
transplanckian physics discussed by various authors (see 
\cite{Danielsson:2005cc,Greene:2005aj,Schalm:2004xg} and references therein).
The barrel instanton background is simply deSitter plus radiation in
Lorentzian signature metric. Doing a quantum field theoretic calculation
for the scalar field in this background will shed light on possible
changes in CMB spectrum. This is work is progress \cite{Sarangi:XX}.
It is, in fact, very interesting to note that a source term arises very
naturally when a physical (not comoving) momentum cut-off is imposed
on a de Sitter background \cite{Danielsson:2005cc,Keski-Vakkuri:2003vj,
Brandenberger:2004kx}. The radiation term that leads to the modified
wavefunction and that depends on the cut-off might be seen as such a
source term. This possibility deserves further investigation. \\

It is natural question to ask what sense there is to calculating 
probabilities when one is comparing different universes. 
Everett and Wheeler \cite{Everett:1957hd} were led to their formulation
of this problem and a possible solution while asking these questions.
If the wavefunction of the universe is to be used to calculate
probabilities that involve different universes on a stringy landscape, 
then one is 
naturally confronted by these issues. Further work is required 
along the directions set by \cite{Everett:1957hd}. \\

 Finally, we would like to address the criticism of the modification
of the wavefunction mentioned in \cite{Holman:2005eu}. The authors
raised the possibility of interpreting the radiation term simply
as a renormalization of the cosmological constant due to the
vacuum energies of the perturbative modes. This, however, is not
true. First, a string scale appears directly in the correction term, thus introducing a 
new scale into the tunneling formula. 
As we have discussed in Sec.($4.1$), the perturbative modes
not only renormalize the constant, but also introduce a radiation term
due to the quantum fluctuations of the metric inherent in a 
de Sitter spacetime. The fluctuations of the background lead
to the radiation term that cannot be absorbed into any counter-term.
QFT in curved background assumes a fixed background and the matter
field is a higher order correction to it. This enables one to do
a semiclassical treatment and absorb away the cut-off dependent 
quantities. However, having claimed that when the background metric
itself is fluctuating, as in a de Sitter spacetime, there will
be a source term.

 \vspace{0.5cm}

{\large{\bf{Acknowledgments}}}\\

We thank Andrei Linde for vigorous discussions that motivated this note.
We also thank Faisal Ahmad, Andrei Barvinski, Brian Greene, Louis Leblond,
Koenraad Schalm,  Jan Pieter van der Schaar, Sarah Shandera, Gary Shiu, 
Ben Shlaer and Erick Weinberg for useful discussions. 
This work is supported by the National Science Foundation under 
Grant No. PHY-009831 and by the DOE under Grant No. DE-FG02-92ER40699.

\end{document}